\documentclass[conference, 10pt]{IEEEtran}
\makeatletter
\def\ps@headings{%
\def\@oddhead{\mbox{}\scriptsize\rightmark \hfil \thepage}%
\def\@evenhead{\scriptsize\thepage \hfil \leftmark\mbox{}}%
\def\@oddfoot{}%
\def\@evenfoot{}}
\makeatother
\usepackage{times}
\usepackage[dvips]{graphicx}

\usepackage{euler}
\usepackage{amsmath,amssymb,amsthm}
\usepackage{epsfig,subfigure}
\usepackage[section]{algorithm}
\usepackage{algorithmic}
\usepackage{comment}
\usepackage{cite}
\usepackage{geometry}
\usepackage{color}
\usepackage{subfigure}
\usepackage{epstopdf}

\renewcommand{\baselinestretch}{1.00}

\usepackage{geometry}
\usepackage{epstopdf}

\begin{document}
%
\title{Location Aided Energy Balancing Strategy in Green Cellular Networks}


\author{\IEEEauthorblockN{Jianjun Yang}
\IEEEauthorblockA{Department of Computer Science\\
University of North Georgia\\
Email: jianjun.yang@ung.edu}

\and
\IEEEauthorblockN{Bryson Payne}
\IEEEauthorblockA{Department of Computer Science\\
University of North Georgia, USA\\
Email:  bryson.payne@ung.edu}

\and
\IEEEauthorblockN{Markus Hitz}
\IEEEauthorblockA{Department of Computer Science\\
University of North Georgia, USA\\
Email: markus.hitz@ung.edu}

\and
\IEEEauthorblockN{~~~~~~~~~~~~~Zongming Fei}
\IEEEauthorblockA{~~~~~~~~~~~~~Lab for Advanced Networking\\~~~~~~~~~~~~~Department of Computer Science \\
~~~~~~~~~~~~~University of Kentucky\\
~~~~~~~~~~~~~Email: fei@netlab.uky.edu}

\and
\IEEEauthorblockN{Le Li}
\IEEEauthorblockA{David R.heriton School \\of Computer Science \\
University of Waterloo, Canada\\
Email: l248li@uwaterloo.ca}

\and
\IEEEauthorblockN{Tongquan Wei}
\IEEEauthorblockA{School of Information Science \\and Technology \\
East China Normal University, China\\
Email:  tqwei@cs.ecnu.edu.cn}

}


%


\maketitle

\begin{abstract}
Most cellular network communication strategies are
focused on data traffic scenarios rather than energy balance and efficient utilization.
Thus mobile users(cell phones) in hot cells
may suffer from low throughput due to energy loading imbalance
problem. In state-of-art cellular network technologies, relay stations extend cell coverage and
enhance signal strength for mobile users. However, busy traffic makes the relay stations in hot area run out of energy quickly.
In this paper, we propose an energy balancing strategy in which the mobile nodes are able to dynamically select and hand over to the
relay station with the highest potential energy capacity  to resume communication.
Key to the strategy is that each relay station merely maintains two  parameters that contains the trend of its
 previous energy  consumption and then predicts its
future quantity of energy, which is defined as the relay station's  potential energy capacity. Then each  mobile node can select the relay station with the highest potential energy capacity.
 Simulations demonstrate that our
approach significantly increase the aggregate throughput and the average life time of relay stations in  cellular network environment.

\end{abstract}


%
\IEEEpeerreviewmaketitle

\section{Introduction}
During the past several years, there has been tremendous growth in cellular networks.
With the introduction of Android and iPhone devices, use of ebook readers such as iPad and Kindle and the success of social networking giants such as Facebook, WeChat and QQ,
the number of customers and the demand for cellular traffic have escalated significantly\cite{shenju1}\cite{shenju2} \cite{shenju3}\cite{shenju4}.
The huge amount of customers and very high volume data transmission result in serious problems of energy consumption\cite{cite_1_intro}.

The rising energy costs of operating cellular networks have led to an emerging trend of addressing energy efficiency and energy balance utilization amongst the network operators and regulatory bodies such as 3GPP and ITU\cite{cite_4_intro}. This trend has stimulated the interest of researchers in an innovative new research area called “green cellular networks”. In this regard, the European Commission has recently started new projects
to address the energy issue of mobile communication systems, such as ``Energy Aware Radio and Network Technologies (EARTH)'', ``Towards Real Energy-efficient
Network Design'' and ``Cognitive Radio and Cooperative strategies for Power Saving in Multi-standard Wireless Devices'' \cite{cite_6_intro}. Energy in cellular network  is a vast research discipline that needs to cover all the layers of the protocol stack and various system architectures and it is important to identify the fundamental trade-offs linked with energy efficiency and the overall performance \cite{cite_9_intro}.

Scholars have addressed four key issues in terms  of energy efficiency with network performance \cite{cite_4_intro}\cite{xukunjie2} . They are deployment efficiency (balancing deployment cost, throughput), spectrum efficiency (balancing achievable rate), bandwidth (balancing the bandwidth utilized) and delay (balancing average end-to-end service delay). To address the challenge of increasing power efficiency in future cellular networks and thereby to maintain profitability, it is crucial to consider various paradigm-shifting technologies, such as energy efficient wireless architectures and protocols, efficient base station(BS) redesign, smart grids, opportunistic network access or cognitive radio, cooperative relaying and heterogeneous network deployment based on smaller cells.

In addition to academia, governments and industries have recently shown keen concerns on the critical issues related to energy efficiency and its balance utilization and  in the ICT(Information and Communication Technology) area. However, as studied in the recent literature \cite{cite_3_intro}\cite{cite_92_intro}, most of the techniques applied to current mobile networks have been designed by taking into account non-energy-related factors, such as throughput, Quality of Service (QoS), availability, scalability, and so on\cite{qsun1}\cite{qsun2}\cite{qsun3}\cite{zhang2012gossip}\cite{zhang2013capacity}\cite{zhang2013mobile}\cite{junew1}. Moreover,  in real-world systems,
mobile users (MUs) are not evenly distributed across cells, resulting in that MUs in a hot cell will be affected by the load imbalance and they might unable
to get services.
Meanwhile, as novel network architectures that include picocells, hierarchical cells and femtocells emerge,
the density of base stations and mobile users is becoming larger, and the cells go smaller.
The appearance of this high density cellular networks introduces more variety of load across different cells and makes the load imbalance problem more serious\cite{ywang1}\cite{ywang3}.

The new generation cellular networks allow mobile users to connect relay station and then connect to base station.  The busy traffic near a relay station often makes the energy of the relay station go down very quickly. However, the relay station with low traffic is idle with high energy.
In order to balance the energy utilization among different cells, it is needed to transfer the over-loaded traffic from hot cells to neighboring cooler ones.
The challenge is how to balance the energy utilization of the relay stations in order to get the best trade-off among all the relay stations in a cellular network.

In this paper, we develop a statistical parameter  based energy balance utilization algorithm, in which each relay station maintains the
acceleration and the variance of energy consumption  that  represent the station's historical energy consumption acceleration and the variance. Together with its current energy quantity,
the relay station is able to predict its future energy quantity, which is considered as  its potential energy capacity.
Before payload data transmission between a mobile user and a relay station, the mobile user disseminates a message to a portion of
its neighbor relay stations depending on their geographic locations . Then it selects and hands over to the
  relay station with highest potential energy.  Because each relay station only needs to maintain
two parameters, the overhead of this scheme is very low. Our simulation results
illustrate that our approach  significantly increases the aggregate throughput in the network and the average life time of the relay stations
compared
with existing approaches.

The rest of the paper is organized as follows. Section II discusses the related research on this topic. Section III proposes
a novel method that select the best relay station. We evaluate the proposed schemes by simulations
and describe the performance results in Section IV. Section V
concludes the paper.

\section{Related Work}

Various energy utilization strategies in cellular networks were proposed and scholars developed various ways to solve energy utilization questions\cite{leli4}\cite{lile1}\cite{lile2}\cite{lile3}. The schemes are classified  into several categories. The first one contains  the strategies based on channel borrowing from cooler cells\cite{cite_1_related}. The second category includes the strategies  based on BS selection\cite{cite_3_related}. The third category contains  strategies based on power control and cell breathing\cite{cite_3_related}\cite{cite_4_related}.
The last  type consists of strategies on relay-assisted traffic transfer\cite{cite_7_related}. The basic idea of channel borrowing is to borrow a set of channels from cooler cells to hot cells.
However, this will change the pre-defined spectrum reuse pattern and introduce more co-channel interference\cite{yangchannel}.
There are also a great deal of research combining multi-hop wireless network (MWN) with infrastructure wireless networks, forming architecture of multi-hop
cellular network (MCN). In MCNs, relay stations (RSs) are network components that are dedicated to storing and forwarding data received from BSs to MUs, and vice versa.
Deploying relay stations can extend the coverage and enhance the signal strength, which clearly help improve the performance for MUs near the edge of the cell. The appearance of MCNs also provides another method to solve the load imbalance among different cells. That is to transfer over-loaded traffic from hot cells to cooler cells by RSs. Compared with previous discussed dynamic load balancing schemes, relay-based load balancing schemes are more flexible and will introduce less interference. In\cite{cite_3_intro}, a mobile-assisted call admission scheme is proposed to achieve load balancing in cellular networks, which require an ad-hoc overlay network on the cellular network. The authors divided the channels into two parts, one for the ad-hoc overlay network, the other for the cellular network.
In\cite{cite_92_intro}\cite{yangchannel}, the authors proposed dynamic load balancing schemes in the integrated cellular and ad-hoc relaying systems (iCAR). The ad-hoc relaying stations (ARS) compose an overlay ad-hoc network, which can help relay traffic among different cells. As in iCAR systems, ARSs work on ISM-band channels, it is pointed out that the performance of iCAR systems will depend on the number of available ISM-band channels, and interference in ISM-band could affect the performance of dynamic load balancing.

A. Alam et al. \cite{cite_Alam} investigates dynamic traffic-aware BS switching modes, where the BS can alter its operating modes between standard BS operations and switching to relay station (RS) mode, the so-called BS-RS Switching model. Depending on the traffic fluctuations, load profiles are divided into two categories.
One is Zero-to-medium traffic period  when a BS switches to the RS mode and turns off all its high-power consuming equipment. The second is Peak traffic period when all BS are fully active. The rationale for switching from a BS to RS mode is to ensure those MS that would be served by the switched off cell and may suffer deep fading, are still be able to receive the same QoS. Furthermore, since the propagation distance has been shortened between the MS and serving BS via the back haul connection, the required MS transmit power is concomitantly reduced compared with the BS sleep and cell zooming technique [5].

K. Xu and M. Zhou \cite{cite_KunXu1}\cite{cite_KunXu2} developed a realistic and representative energy model based on RF transceiver. As a typical IEEE 802.15.4 RF transceiver mode, each wireless node is equipped with multiple discrete transmit power levels. Based on this energy model, they developed a power control policy, which configure the transmit power level as a function of transmission distance such that the energy cost is minimized. Then, the optimized energy balanced chain model is proposed to determine the optimal traffic flow distribution.


\section{Problem Formulation}
\label{mainalg}
\subsection{The Basic Idea}
\begin{figure}[!htp]
\begin{center}
\includegraphics[width=8.0cm]{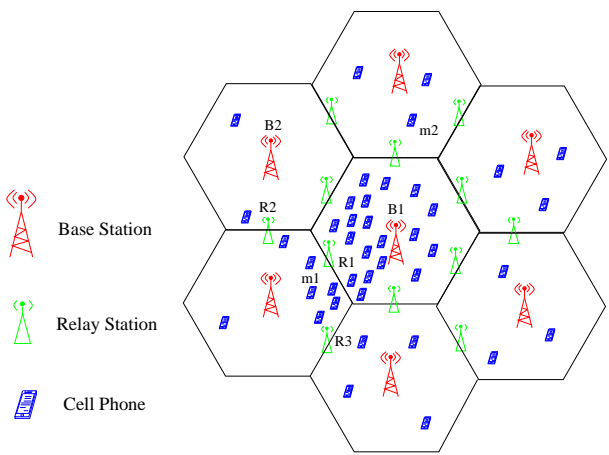}
\end{center}
\caption{A topology of a cellular network.}
\label{fig1}
\end{figure}

In cellular networks, each base station covers a number of cell phones in a hexagonal cell. We consider a cellular network containing a set of base stations(BS), relay stations(RS) and mobile users(cell phones)(MU), denoted by $\left\{B_1, B_2,...,B_s\right\}$,
$\left\{R_1, R_2,...,B_t\right\}$, and $\left\{m_1, m_2,...,m_p\right\}$.  Each BS is located at the center of the cell. Each RS is located at a boundary of cells.
The traditional communication between two mobile users $m_i$ to $m_j$ is that $m_i$ communicates to its base station $B_i$, then $B_i$ communicates to $B_j$ which covers
$m_j$, and then $B_j$ communicates to $m_j$. In new generation cellular network, the relay stations are used to assist communications between
cell phone to cell phone or cell phone to base station.
Fig.~\ref{fig1} shows an instance with several base stations, relay stations and cell phones, where the central cell is a hot cell with a large number of cell phones and busy data traffic and the other
cells are cooler cells with less cell phones. Note that the cell phones in all figures of this paper represent active cell phones.

Data transmission takes place when a cell phone(initiator) intends to talk to another cellphone(receiver). The scenario is simple if the two are in the same
cell\cite{thesisJosh}. We consider the circumstance
that the receiver is in another cell.  In Fig.~\ref{fig1}, the communication between $m_1$ and  $m_2$ is such an instance, in which data is relayed from $m_1$ to $R_1$, then from $R_1$ to $B_1$, and continues until it reaches $m_2$. In the scenario, $R_1$ is hot with a lot of cell phones nearby. Hence its energy is going down very quickly. However, $R_2$ covers less cell phones and its energy is going down slowly. Here comes up with a question, can $m_1$ find $R_2$ or even $R_3$ to relay the message  to bypass the hot relay station $R_1$?

\begin{figure}[!htp]
\begin{center}
\includegraphics[width=5.0cm]{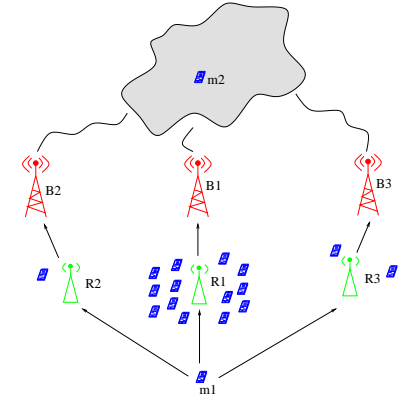}
\end{center}
\caption{The scenario to illustrate the main idea.}
\label{fig2}
\end{figure}

Since a mobile cell phone is able to select a relay station and hand over to it\cite{{thesisJosh}}, we develop an approach to let each mobile user find the
RS with highest potential energy and hand over to it. We start our basic idea by a simplified case as Fig.~\ref{fig2}. Suppose $m_1$ intends to talk to $m_2$. If  $m_1$ selects  $R_1$ to conduct relay,
it may not be a good choice because $R_1$ is busy and may run out of energy shortly.
$R_1$, $R_2$ and $R_3$ are all possible relay stations for $m_1$.
Suppose the residual energy of $R_1$ is 900 $kWh$, $R_2$ is 800 $kWh$ and $R_3$ is 850 $kWh$ at current moment.
We further assume that the historical residual energy vector up to this moment of
 $R_1$ is [..., 2000, 1500, 1300, 900],    $R_2$ is [..., 900, 870, 830, 800], and   $R_3$ is [..., 1400, 1200, 1000, 850].
A simple strategy will let $m_1$ select the one that has the highest energy at the moment.
Seemingly, $R_1$ is the best choice. However, since $R_1$ is in a hot area  serving a large number of mobile users and its energy is going down dramatically, while $R_2$
are serving much less mobile users and its energy goes down slowly, $m_1$  should select $R_2$.

Our goal is to let $m_1$ select the relay station that will have the highest potential energy. In our approach,
each RS can predict its future energy right after this time slot based on its historical and current energy, which is defined as  potential energy of the RS
in this paper.
In this example, our strategy will let $m_1$ choose $R_2$.

\subsection{Computation of Potential Energy}

The mobile initiator  needs to know which relay station is the one that has the potential highest  energy.
The energy varying trend  is essential to determine an RS's potential   energy.
Suppose the current time is $k$. A naive method on predicting the energy at time $k+1$ is as follows. Let RS $R_i$ save
all its historical and current (time $k$) energy $E_{i_0}$, $E_{i_1}$, ...,$E_{i_k}$  in a vector.
Then it can predict its energy for time $k+1$ by  curve-fitting in numerical analysis that approximates
its moving trend. The limitations of this method are to spend too much space to save the energy values in the vector and too complex for curve-fitting computation.

We propose a statistical based strategy to let each relay station predict  its future energy. In this scheme, each RS only needs to maintain two parameters. One is its potential acceleration and the second is the
variance of the acceleration. The two parameters store how its residual energy changes. Together with the RS's current energy,  its future  energy capacity can be calculated
and is considered its potential energy.
Thus, mobile user $m_1$ is able to figure out the best RS subsequently.

We define the following notations to represent the terms regarding our approach for node $R_i$.

\noindent $E_{i_k}$: The measured energy at time $k$. \\
 $\hat{E_{i_k}}$: The energy predicted at time $k$. \\
$a_{i_k}$: The acceleration measured  at time $k$. It indicates how energy changes during a time slot.\\
$\hat{a_{i_k}^-}$: The acceleration at time $k$ evolved   from time $k-1$.\\
$\hat{a_{i_k}}$: The potential acceleration at time $k$. \\
 $v_{i_k}$: The variance of acceleration updated  at time $k$.\\
  $v_{i_k}^-$: The variance of acceleration at time $k$   evolved from time $k-1$.\\
$\epsilon$: The error or noise in the process.\\
  $B_{i_k}$: The blending factor at time $k$.\\

At time $k$, $R_i$  measures its energy $E_{i_k}$. And then it computes $a_{i_k}$ as its measured acceleration  by

\[a_{i_k}=(E_{i_k}-E_{i_{k-1}})/\Delta t   ~~~(1)\]

$R_i$ updates $\hat{a_{i_k}^-}$ and $v_{i_k}^-$ in order to keep its historical energy to predict its future energy.
\[\hat{a_{i_k}^-}=\hat{a_{i_{k-1}}} ~~~(2)\]
\[v_{i_k}^-=v_{{i_k-1}}+\epsilon  ~~~(3)\]
$R_i$ also computes the blending factor $B_{i_k}$,
which indicates how much the acceleration changes from last time to current time.\\
\[B_{i_k}=v_{i_k}^-(v_{i_k}^-+\epsilon)^{-1}=v_{i_k}^-/(v_{i_k}^-+\epsilon) ~~~(4)\]
Once $R_i$ obtains the blending factor $B_{i_k}$ and the evolved acceleration $\hat{a_{i_k}^-}$, it knows how much the acceleration changes and the
evolved acceleration. Additionally, $R_i$ considers the measured acceleration $a_{i_k}$. Then it calculates its potential acceleration
$\hat{a_{i_k}}$. This acceleration will be used to predict its energy of time $k+1$.
\[\hat{a_{i_k}}=\hat{a_{i_k}^-}+B_{i_k}(a_{i_k}-\hat{a_{i_k}^-}) ~~~(5)\]
$R_i$ updates the variance of acceleration for future utilization.
\[v_{i_k}=(1-B_{i_k})v_{i_k}^- ~~~(6)\]
Finally, the energy predicted of time $k+1$ is
\[E_{i_{k+1}}=E_{i_k}+\hat{a_{i_k}}\Delta t ~~~(7)\]
At a certain time $k$, $R_i$ only needs to measure its energy $E_{i_k}$ and record two parameters
$\hat{a_{i_{k-1}}^-}$ and $v_{i_{k-1}}$. Then it can predict its energy at time $k+1$ by the calculation, which is the potential energy capacity of $R_i$.

\subsection{Algorithm Description}

\begin{figure}[!htp]
\begin{center}
\includegraphics[width=5.0cm]{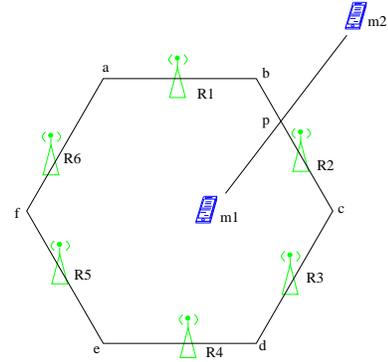}
\end{center}
\caption{To figure out three good relay stations.}
\label{fig3}
\end{figure}

In our approach, when a user $m_1$ intends to communicate to $m_2$, it selects the relay station with highest potential energy.
Apparently, $m_1$ does not need to consider the relay stations in the opposite direction  $m_1$  $\rightarrow$  $m_2$,  such as $R_4$, $R_5$ and $R_6$ in Fig.~\ref{fig3}.
Here $R_1$, $R_2$ and $R_3$ are  candidates for $m_1$. How can  $m_1$ know the three possible candidates  among the six relay stations in the boundaries? In new generation cellular networks, each device is equipped with
GPS and hence it knows  its location. We assume that the initiator $m_1$ knows the location of the receiver $m_2$. The assumption is very common in geographic routing\cite{yangex1}\cite{yangex2}\cite{yangex3}. So $m_1$ finds out the three possible candidates in this way: It connects $m_1$ and $m_2$. The line segment intersects  the hexagonal area at $p$. Then the relay station located at the edge where $p$ is located  and the two relay stations located at its adjacent edges
are considered as good relay candidates. In Fig.~\ref{fig3}, $p$ is located at segment $bc$, so $R_2$ located at $bc$ is considered as one of the candidates.
The two adjacent edges of $bc$ are $ab$ and $cd$, so $R_1$ and $R_3$ are also considered as $m_1$'s candidates.

In our strategy, when a cell phone(initiator) intends to talk to a receiver in a different cell, it  calls the procedure Detection() to figure out the three possible relay stations by its location
and the destination's location. In the procedure, it sends out a detection message to the three relay stations. Upon receiving the detection message, each RS calls
Prediction() to calculate its potential
energy  and inform the energy capacity to the initiator. The initiator calls Selection() to find the relay station with possible highest potential energy,
hands over to it and sends its payload to the RS.


\begin{figure}[hbt]
\baselineskip=10.2pt
{\small
  \noindent\underline{\bf Detection()} \\
  \text{~~~~} Figure out the three candidate relay stations by its location and destination's location\\
  \text{~~~~} Send detection message to the three relay stations
  }
\protect\caption{Algorithm for initiator(MU) to send detection message}
\protect\label{routingfig}
\end{figure}

\begin{figure}[hbt]
\baselineskip=10.2pt
{\small
  \noindent\underline{\bf Prediction()} \\
  \text{~~~~} Calculate its potential energy by formula (7)\\
  \text{~~~~} Send the energy quantity back to the initiator
  }
\protect\caption{Algorithm of computation for potential energy}
\protect\label{routingfig}
\end{figure}

\begin{figure}[hbt]
\baselineskip=10.2pt
{\small
  \noindent\underline{\bf Selection()} \\
  \text{~~~~} Compare the three energy quantities from the three relay stations\\
  \text{~~~~} Select the RS with highest energy values\\
  \text{~~~~} Hands over to the RS\\
  \text{~~~~} Send payload to the RS
    }
\protect\caption{Algorithm of hand over}
\protect\label{routingfig}
\end{figure}

\section{Performance Evaluation}

The performance evaluation was conducted in a simulated noiseless
radio network environment using MATLAB. We create a topology that consists of 20 hexagonal cells. A base station is located in the center of
each cell  and a relay station is located at each edge. We randomly distribute a number of mobile users to the cells.
We performed a sequence of experiments in which the number of mobile users is changed from 200 to 800 with increment of 50. For each number of mobile users, we measure the aggregate throughput  10 times
and present the average. We also measure the average life time of the relay stations in our experiments.
The  life time of a relay station is important. If one RS runs out of energy and turns down, there will be a hole\cite{yangitgr} and the whole throughput will be
decreased dramatically.

Our approach considers energy balance. We compare our approach(``EB by MU") with two other approaches. One is the scheme which
considers  energy balance but the balance is determined by base station(``EB by BS"). The second one is without considering energy balance(``No EB").

\begin{figure}[!htp]
\begin{center}
\includegraphics[width=8.0cm]{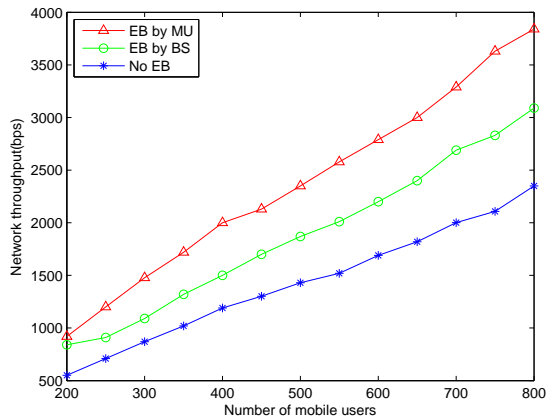}
\end{center}
\caption{Aggregate throughput} \label{g22}
\end{figure}

Fig.~\ref{g22} shows our evaluation of aggregate throughput. 
Among the three approaches, the scheme without energy balance(``No EB") has the lowest throughput. This indicates that energy balance can improve aggregate throughput
because the traffic are allocated more reasonable to the relay stations. The  throughput  of our method is the highest one.

\begin{figure}[!htp]
\begin{center}
\includegraphics[width=8.0cm]{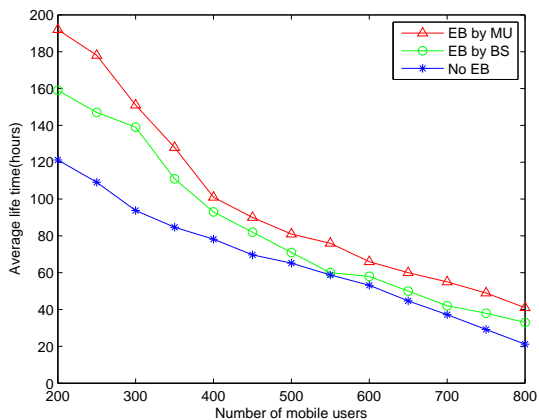}
\end{center}
\caption{Life time of relay stations}
\label{g11}
\end{figure}

Fig.~\ref{g11} shows the life time of the three schemes. The one without energy balance(``No EB") results in the lowest average lift time. That is because when a relay station has busy traffic load, it runs of out energy quickly. And then there is a void or hole. The regular solution for hole problem is to find the landmark relay station
\cite{yangitgr}.
However, the landmark has very heavy burden and it is exhausted quickly. This makes the network much worse. Our approach
has the longest lift time because the communication between two cell phones is always relayed by the RS with the highest potential energy. If a RS is busy and its energy goes down
significantly, the new communication will not rely on it. Hence the balanced energy utilization makes the life time longer.

\section{Conclusion}
\label{conclusion}

In this paper, we present a statistical based mechanism to balance energy utilization in cellular network.
In the approach, each relay station only needs to remain two factor to store its
historical energy trend and then it is able to predict its future energy. The message initiator selects the
relay station with
potential highest  energy to relay data transmission.
Simulations demonstrate that our approach results in higher throughput in the network and longer average life time of relay stations over related approaches.



%

\end{document}